\documentclass[prd,nofootinbib,floatfix,preprintnumbers, twocolumn, superscriptaddress, showpacs]{revtex4-1}%
\pdfoutput=1 

\usepackage{amssymb,amsmath}
\usepackage{color}
\usepackage{graphicx}
\usepackage{comment}
\usepackage{epsfig}
\usepackage{latexsym} % Provided \Box\begin{figure}[htbp]
\usepackage{multirow}

\def\beq{\begin{equation}}
\def\eeq{\end{equation}}
\def\barr{\begin{eqnarray}}
\def\earr{\end{eqnarray}}
\def\bc{\begin{center}}
\def\ec{\end{center}}

\newcommand{\osq}{\frac{1}{\sqrt{2}}}

\begin{document}

\title{Hybrid Baryons in QCD}

\author{Jozef~J.~Dudek}
\email{dudek@jlab.org}
\affiliation{Jefferson Laboratory, 12000 Jefferson Avenue,  Newport News, VA 23606, USA}
\affiliation{Department of Physics, Old Dominion University, Norfolk, VA 23529, USA}

\author{Robert~G.~Edwards}
\email{edwards@jlab.org}
\affiliation{Jefferson Laboratory, 12000 Jefferson Avenue,  Newport News, VA 23606, USA}

\collaboration{for the Hadron Spectrum Collaboration}
\date{January 10, 2012}

\begin{abstract}

We present the first comprehensive study of hybrid baryons using lattice QCD methods. Using a large basis of composite QCD interpolating fields we extract an extensive spectrum of baryon states and isolate those of hybrid character using their relatively large overlap onto operators which sample gluonic excitations.

We consider the spectrum of Nucleon and Delta states at several quark masses finding a set of positive parity hybrid baryons with quantum numbers $N_{1/2^+},\,N_{1/2^+},\,N_{3/2^+},\, N_{3/2^+},\,N_{5/2^+},\,$ and $\Delta_{1/2^+},\, \Delta_{3/2^+}$ at an energy scale above the first band of `conventional' excited positive parity baryons. This pattern of states is compatible with a color octet gluonic excitation having $J^{P}=1^{+}$ as previously reported in the hybrid meson sector and with a comparable energy scale for the excitation, suggesting a common bound-state construction for hybrid mesons and baryons.

\end{abstract}

\preprint{JLAB-THY-12-1479}

\maketitle

\section{Introduction}\label{sec:intro}
While QCD is the accepted underlying theory of hadronic physics, with hadrons being viewed as bound states of strongly coupled quarks and gluons, the role of excited gluonic fields in determining the spectrum of mesons and baryons remains unclear. QCD in the strongly coupled low-energy regime gives rise to a number of interesting phenomena that are exhibited in the spectrum of mesons and baryons, such as the spontaneous breaking of chiral symmetry yielding effective `constituent' quark degrees-of-freedom. These quasi-particles have the quantum numbers of quarks, and there are clear signs of a spectrum of excitations corresponding to their relative motion. It seems odd then, considering that the gluonic field is strongly coupled to itself and to quarks, that there are no unambiguous signals for states featuring a gluonic excitation.

The simplest place to look for gluonic excitations would seem to be in the isoscalar meson spectrum, where states built purely out of glue, the `glueballs', could be present. This has proven to be very difficult in practice \cite{Crede:2008vw} with suggestions that $q\bar{q}$ excitations mix strongly with glueball basis states. Another target is elsewhere in the spectrum of mesons, where states built from $q\bar{q}$ augmented by a gluonic excitation can have $J^{PC}$ quantum numbers not available to a pure $q\bar{q}$ state. These `exotic hybrid mesons' remain our best hope of a `smoking gun' experimental signature, but in practice the current experimental situation is at best confused (see the review in \cite{Meyer:2010ku}). Theoretical understanding of hybrid meson states in QCD recently took a significant step forward with the application of lattice methods to the problem. In \cite{Dudek:2009qf,Dudek:2010wm,Dudek:2011tt,Dudek:2011bn} the spectrum of mesons extracted from lattice QCD computations was interpreted in terms of `conventional' $q\bar{q}$ states supplemented with a spectrum of hybrid mesons, some with exotic $J^{PC}$ and some with conventional $J^{PC}$. The degeneracy pattern of these states and the form of the composite QCD operators used to interpolate them from the vacuum strongly suggested that the gluonic excitation present in hybrid mesons is of chromomagnetic character.

Hybrid baryons have not attracted the same attention as hybrid mesons principally because they lack manifest `exotic' character. All $J^P$ values can be populated by states constructed from three quarks having excitation in orbital angular momentum, so that hybrids can only appear in terms of over-population with respect to some model of $qqq$ excitations. While a dynamical model may suggest peculiar decay characteristics for hybrid baryon states that might help to identify them, there is not the `smoking gun' quantum number signal present in the meson sector. Furthermore the current experimental situation, in which far fewer baryon resonances are observed than are expected in $qqq$ bound-state models, does not encourage adding \emph{additional} gluonic degrees-of-freedom to the bound-state system. 

Despite there being, as yet, no simple way to extract them experimentally, there is significant theoretical insight to be gained from studying hybrid baryons, particularly in a framework that can simultaneously calculate the properties of hybrid mesons. We might imagine, given the different number of quarks, and the corresponding differing distributions of color sources, that the spectrum of gluonic excitations in a baryon could be different to that in a meson. We will explore this possibility in this paper.

The spectrum of hybrid baryons has previously been considered in a number of QCD-motivated models. In the bag model \cite{Barnes:1982fj}, in which quark and gluon fields are confined within a cavity with the fields satisfying appropriate boundary conditions at the wall of the cavity, a TE-mode gluonic field (transforming as $J^{PC}=1^{+-}$) combines with three quarks in an overall color octet to produce a lightest set of hybrid baryons : 
\begin{align}N_{1/2^+}%{\tfrac{1}{2}^+}
, \,N_{1/2^+}, \, N_{3/2^+}, \,N_{3/2^+}, \, N_{5/2^+} \nonumber \\ 
\Delta_{1/2^+},\,\Delta_{3/2^+}. \nonumber
\end{align}
The overall mass scale of these states is somewhat plastic, depending upon how one chooses to set parameters within the model - in \cite{Barnes:1982fj}, the first excited pion resonance, the $\pi(1300)$, was assumed to be a hybrid meson which then sets the scale for Nucleon hybrids starting as light as 1.6 GeV. A more plausible modern candidate for a hybrid pion would be the second excited pion resonance, the $\pi(1800)$, which would place the hybrid Nucleons significantly heavier, somewhere above 2 GeV.

The QCD sum-rule study in \cite{Kisslinger:1995yw} considered the possibility of a hybrid baryon as the first-excited state above the nucleon with $N_{1/2^+}$ quantum numbers concluding that such a state would lie close to 1.5 GeV. 

An implementation of the flux-tube model\cite{Isgur:1983wj,Isgur:1984bm}, in which quarks sit on the ends of strings, providing a linear confining potential, suggests a set of lightest hybrid baryons~\cite{Capstick:2002wm}:
\begin{align}N_{1/2^+}, \,N_{1/2^+}, \, N_{3/2^+}, \,N_{3/2^+} \nonumber \\ 
\Delta_{1/2^+},\,\Delta_{3/2^+},\, \Delta_{5/2^+}. \nonumber
\end{align}
This differs from the bag model spectrum by having the $J^P=5/2^+$ state with isospin-3/2 rather than isospin-1/2. The mass scale of these states is found to be around 1.9 GeV. The flux-tube model is an example of a framework where we would not necessarily expect there to be a common energy scale for the gluonic excitation in a baryon and in a meson. A meson contains a single flux-tube between the quark and anti-quark which in a hybrid is excited in transverse oscillation. In a baryon however there are three tubes which either meet at a junction or form a triangle. There would seem to be no particular reason the excitations of this more complicated system need be the same as a single tube.

As mentioned earlier, a major challenge for the experimental isolation of hybrid baryons is that they have non-exotic quantum numbers and hence appear embedded in a spectrum of conventional baryon states. In fact, worse than this there is, a priori, nothing within QCD to prevent hybrids mixing arbitrarily strongly with conventional $qqq$ states of the same $J^P$ to produced mass eigenstates which are neither one nor the other. This complication will also appear in any sufficiently sophisticated theoretical approach. Whether QCD in fact manifests strong mixing will need to be determined in explicit calculation.

In this paper we will extract a spectrum of QCD eigenstates from baryon correlators computed using discretisation of QCD on a finite lattice. This technique is an {\it ab-initio} non-perturbative method starting from a gluon and quark lagrangian that allows us to compute the spectrum directly from a controlled approximation to QCD. In order to explore the spectrum of states with a given set of quantum numbers it has proved effective \cite{Michael:1985ne,Luscher:1990ck,Blossier:2009kd, Dudek:2009qf,Dudek:2010wm,Dudek:2011tt}
to construct a large basis of hadron interpolating fields, featuring combinations of quark and gluon fields having the desired external quantum numbers. A matrix of two-point correlation functions can then be evaluated and by diagonalisation, a variational best estimate of the spectrum of states determined. The eigenvectors obtained in this procedure indicate the optimum linear combination of interpolating fields for each state in the spectrum. 
Quark-gluon bound-state structure interpretational information follows if a particular state has a large overlap with certain characteristic interpolating fields, and large overlap onto multiple interpolating fields of different characteristic structure might suggest large mixing of basis states within QCD.

Our approach will mirror that used to determine a spectrum of hybrid mesons in \cite{Dudek:2009qf,Dudek:2010wm,Dudek:2011bn}, where we found that the lightest supermultiplet of hybrid mesons consists of an exotic $J^{PC}=1^{-+}$ state and three non-exotic states with $J^{PC}=0^{-+},\, 2^{-+},\, 1^{--}$ which are embedded in a spectrum of $q\bar{q}$ excitations. The character of gluonic excitation that appears to give rise to this spectrum is chromomagnetic, having $J^{PC}=1^{+-}$. Heavier positive parity hybrid mesons were also identified that appear to correspond to quark orbital excitation on top of the same chromomagnetic gluonic excitation with states featuring any other type of gluonic excitation apparently significantly heavier.

The corresponding procedure for baryons will be followed in this paper: using a large basis of interpolating fields, including a number which rely upon the presence of a non-trivial gluonic field, we will extract a spectrum of baryon states. By considering the relative size of overlap of the extracted states onto interpolators of characteristically $qqq$ and hybrid form we will identify those which we believe to be hybrid baryons.

As with all lattice QCD calculations, we are somewhat limited in scope by computational restrictions which prevent us from calculating with quarks whose masses are as light as those in the real world. In place of this we will compute with a range of heavier quark masses and study the trend as the quark mass is reduced. Since the computational cost increases with the volume of space-time considered we limit ourselves to a small region a few fermi across discretised with grid points a fraction of a fermi apart.
 
In this work we will report on the first systematic studies of hybrid baryons using lattice QCD methods. We will work with three flavours of dynamical quarks - the strange quark being fixed at its physical mass and the degenerate up and down quarks being varied down from the strange quark mass (where we would have an exact $SU(3)_F$ symmetry) to a value where the pion mass is 400 MeV. This is an extension of the work presented in \cite{Edwards:2011jj} which computed the baryon spectrum on the same lattices, but excluded baryon interpolators featuring a gluonic excitation and which made no observations regarding hybrid baryons.

%%%%%%%%%%%%%%%%%%%%%%%%%%%%%%%%%%%%%%%%%%%%%%%%
%%%%%				OPERATORS				%%%%%
%%%%%%%%%%%%%%%%%%%%%%%%%%%%%%%%%%%%%%%%%%%%%%%%

\section{Baryon Interpolators \& their interpretation}\label{sec:operators}

The construction of a basis of baryon operators respecting the required permutational symmetry of three quark fields and featuring up to two gauge-covariant derivatives was described in \cite{Edwards:2011jj}. This leads to a rather large basis of operators that covers all spins and both parities up to $J^P=\tfrac{7}{2}^\pm$. For lower spins there are sufficiently many operators that variational analysis can lead to extraction of a significant number of excited states.

The baryon interpolators used can be expressed in a compact notation which exposes the permutational symmetries, $\mathsf{\Sigma_{F,S,D}} = \mathsf{S,M,A}$,  
\begin{equation}
\Big( B_\mathsf{\Sigma_F} \otimes \left(S^{P_S}\right)^\mathit{n}_\mathsf{\Sigma_S} \otimes D^{[d]}_{L,\mathsf{\Sigma_D}}  \Big)^{J}, \label{ops}
\end{equation}
where the three factors describe the flavor, Dirac spin and derivative structure of the interpolator. In this report we will consider only baryons of flavor $B_\mathsf{\Sigma_F} = N_\mathsf{M},\, \Delta_\mathsf{S}$, the strangeness-zero members of $\mathbf{8}_F, \,\mathbf{10}_F$ representations, over a range of quark masses. The only exception to this is the special case of three degenerate flavors of quark, i.e. having exact $SU(3)$ flavor symmetry, where in addition to the octet and decuplet states, we'll extract a spectrum of $B_\mathsf{\Sigma_F} = \Lambda_\mathsf{A}$ states, the only member of the $\mathbf{1}_F$ representation. 

The Dirac spin factor, $\left(S^{P_S}\right)^\mathit{n}_\mathsf{\Sigma_S}$, represents the possible combinations of upper and lower components of three Dirac spinors transforming with angular momentum $S$ and parity $P_S$ and having permutational symmetry, $\mathsf{\Sigma_S}$. For some $\left(S^{P_S}\right)_\mathsf{\Sigma_S}$ there are multiple possible constructions and they are labelled by the superscript $\mathit{n}$. 

The derivative factor, 
$D^{[d]}_{L,\mathsf{\Sigma_D}}$, indicates the number of gauge-covariant derivatives applied to the quark fields ($d=0,1,2$) and that they are combined to give objects transforming under rotations with angular momentum $L$ and with permutational symmetry $\mathsf{\Sigma_D}$. 

With a single derivative ($d=1$), excluding a symmetric total derivative, only mixed symmetry\footnote{$\mathsf{MS,MA}$ indicates symmetric or antisymmetric behavior in $1\leftrightarrow2$} possibilities transforming as $L=1$ arise:
\begin{equation*}
  D^{[1]}_{L=1,\mathsf{M}}:\;\left\{
  \begin{array}{l}
  	D_{\mathsf{MS},m}^{[1]} =\frac{1}{\sqrt{6}}\big(2 D^{(3)}_m - D^{(1)}_m - D^{(2)}_m\big) \\
    D_{\mathsf{MA},m}^{[1]} = \osq\big(D^{(1)}_m - D^{(2)}_m\big)
  \end{array}\right.
    \label{eq:deriv}
\end{equation*}
where e.g. $D^{(1)}_m = \vec{\epsilon}(m)\cdot \vec{D}^{(1)}$ is the gauge-covariant derivative, in a circular polarisation basis, acting on the first quark field. 

At the two derivative level ($d=2$) definite permutation symmetry is implemented by the appropriate linear combination of products of $D^{[1]}_\mathsf{MS,MA}$, while the rotational transformational properties are set using an $SO(3)$ Clebsch-Gordan coupling for $1\otimes 1 \to L=0\oplus1\oplus2$:
\begin{widetext}
\begin{align}
  D_{L,\mathsf{S}}^{[2]}  =  \langle 1m;1m'| LM\rangle \osq(D_{\mathsf{MS},m}^{[1]} D_{\mathsf{MS},m'}^{[1]}+D_{\mathsf{MA},m}^{[1]} D_{\mathsf{MA},m'}^{[1]}),\label{eq:2DS}\\
D^{[2]}_{L,\mathsf{M}}:\;\left\{	
 \begin{array}{l}
  D_{L,\mathsf{MS}}^{[2]} = \langle 1m;1m'| LM\rangle \osq(-D_{\mathsf{MS},m}^{[1]} D_{\mathsf{MS},m'}^{[1]}+D_{\mathsf{MA},m}^{[1]} D_{\mathsf{MA},m'}^{[1]}),\\
  D_{L,\mathsf{MA}}^{[2]} =  \langle 1m;1m'| LM\rangle \osq(D_{\mathsf{MS},m}^{[1]} D_{\mathsf{MA},m'}^{[1]}+D_{\mathsf{MA},m}^{[1]} D_{\mathsf{MS},m'}^{[1]})
    \end{array}\right.\label{eq:2DM}\\
D_{L,\mathsf{A}}^{[2]}  = \langle 1m;1m'| LM\rangle  \osq(D_{\mathsf{MS},m}^{[1]} D_{\mathsf{MA},m'}^{[1]} - D_{\mathsf{MA},m}^{[1]} D_{\mathsf{MS},m'}^{[1]}). \label{eq:2DA}
\end{align}
\end{widetext}

In practice, the interpolators (\ref{ops}) correspond to sums of terms of generic structure
\begin{equation*}
	\epsilon_{abc} \Big( D^{n_1} \tfrac{1}{2}(1\pm \gamma^0) \psi \Big)_a \, \Big( D^{n_2} \tfrac{1}{2}(1\pm \gamma^0) \psi \Big)_b \, \Big( D^{n_3} \tfrac{1}{2}(1\pm \gamma^0) \psi \Big)_c
\end{equation*}
where each quark field is projected into either upper or lower Dirac components and is acted upon by either zero, one or two gauge-covariant derivatives. The spin and flavor couplings are suppressed here (details of the construction can be found in \cite{Edwards:2011jj}). The required antisymmetry in color, producing overall color-singlet operators, is implemented using the Levi-Civita symbol. Given this it is clear that the flavor, spin and spatial symmetry representations, $\mathsf{\Sigma_F}, \mathsf{\Sigma_S}, \mathsf{\Sigma_D}$ must be combined to give overall symmetric ($\mathsf{S}$) combinations in order for the interpolator to obey Fermi antisymmetry.

\subsection{``Conventional" $qqq$ interpretation of operators}

A notable subset of the interpolators constructed are those featuring only upper components of Dirac spinors, which we refer to as ``non-relativistic". As shown in Table IV of \cite{Edwards:2011jj}, these operators can be simply classified according to an $SU(3)_F\otimes SU(2)_S \otimes O(3)$ symmetry analogous to what one might have in a (non-relativistic) three-quark bound-state model. In this assignment, the derivatives play the role of orbital angular momentum between the quarks (the $O(3)$ symmetry in most bound-state models).

As a concrete example consider the symmetric construction of two derivatives defined in equation (\ref{eq:2DS}), $D^{[2]}_{L,\mathsf{S}}$, which can be explicitly written
\begin{align}
D^{[2]}_{L,\mathsf{S}}=&\frac{1}{3\sqrt{2}} \big\langle 1m; 1m'\big| LM\big\rangle\nonumber  \\
&\quad\Bigg[ 2\Big(D^{(1)}_{m} D^{(1)}_{m'} + D^{(2)}_{m} D^{(2)}_{m'} + D^{(3)}_{m} D^{(3)}_{m'} \Big)\nonumber \\
 &\quad\quad- \Big( D^{(1)}_{m} D^{(2)}_{m'} + D^{(2)}_{m} D^{(1)}_{m'} \Big) \nonumber \\
 &\quad\quad\quad- \Big( D^{(1)}_{m} D^{(3)}_{m'} + D^{(3)}_{m} D^{(1)}_{m'} \Big) \nonumber \\
 &\quad\quad\quad\quad-  \Big( D^{(3)}_{m} D^{(2)}_{m'} + D^{(2)}_{m} D^{(3)}_{m'} \Big)   \Bigg], \label{D2S}
\end{align}
where expressed more fully a term $D^{(3)}_{m} D^{(2)}_{m'}$ would have the color-spatial structure $\epsilon_{abc}\, \psi_a ( D_{m'} \psi)_b (D_{m} \psi)_c$. A simple model interpretation of the action of $D^{[2]}_{L,\mathsf{S}}$ follows if we treat the gauge-covariant derivatives as ordinary, commuting, derivative operators (neglecting the presence of the gauge-field) and use Jacobi co-ordinates ($\vec{\rho} = \tfrac{1}{\sqrt{2}}(\vec{r}_1-\vec{r}_2),\, \vec{\lambda} = \tfrac{1}{\sqrt{6}}(\vec{r}_1 + \vec{r}_2 - 2\vec{r}_3)$) for the quark positions. In that case we find that 
\begin{align*}
D^{[2]}_{L,\mathsf{S}} &\sim \big\langle 1m; 1m'\big| LM\big\rangle \Big[ (\vec{p}_\rho)_{m} (\vec{p}_\rho)_{m'} + (\vec{p}_\lambda)_{m}(\vec{p}_\lambda)_{m'}\Big] \\
&\sim \big[ p_\rho^2 Y_L^M(\hat{p}_\rho) Y_0^0(\hat{p}_\lambda) + p_\lambda^2 Y_L^M(\hat{p}_\lambda)Y_0^0(\hat{p}_\rho)  \Big]\delta_{L,\mathrm{even}}, 
\end{align*}
so that if $L=2$, we could interpret this as the symmetric combination of a $D$-wave on the $\rho$-``oscillator" coupled to an $S$-wave on the $\lambda$-``oscillator" and a $D$-wave on the $\lambda$-``oscillator" coupled to an $S$-wave on the $\rho$-``oscillator". This kind of operator, coupled to ``non-relativistic" Dirac spinors can interpolate from the vacuum exactly the kind of structures used in $qqq$ ``quark models" \cite{Isgur:1977ef,Isgur:1978xj,Isgur:1978wd,Capstick:1986bm}. For example, the interpolator $\Big(N_\mathsf{M} \otimes \big( \tfrac{1}{2}^+ \big)_\mathsf{M}^\mathit{1} \otimes D^{[2]}_{L=2,\mathsf{S}} \Big)^{\tfrac{3}{2}^+}$ would be identified with the model state $N\,^2\!D_\mathsf{S}\,\tfrac{3}{2}^+$. Table IV in \cite{Edwards:2011jj} presents $qqq$ model interpretations for ``non-relativistic" interpolators.

%%===== HYBRID OPERATORS ======
\subsection{``Hybrid" $qqqG$ interpretation of operators}

The subset of operators which we identify as essentially hybrid in nature appears at two derivatives and follows from combinations which correspond to \emph{commutation} of two gauge-covariant derivatives acting on the same quark field. Such operators are zero in a theory without gauge-fields and correspond to the chromomagnetic components of the gluonic field-strength tensor. This is the central observation that leads us to identify these operators as being `hybrid' - unless an eigenstate contains a gluonic field in a non-trivial configuration it should have no overlap onto an operator of this form.

In the derivative constructions described above such entries appear via terms of the form $\langle 1m;1m'|1M\rangle D^{(n)}_m D^{(n)}_{m'} \propto \epsilon(M)_i \epsilon_{ijk} [D_j,D_k] \propto B_M$, where the chromomagnetic field in a circular polarisation basis, $B_M \equiv \vec{\epsilon}(M)\cdot \vec{B}$ is related to the commutator of two gauge-covariant derivatives, $B_k = \tfrac{1}{2}\epsilon_{ijk} [D_i, D_j]$. As an example of how such an object enters into our operator basis, consider the projection of equations (\ref{eq:2DM}) into $L=1$:
\begin{align*}
D^{[2]}_{L=1,\mathsf{MS}}=&\frac{1}{3\sqrt{2}} \big\langle 1m; 1m'\big| 1M\big\rangle \\
&\quad\Bigg[ 2\Big(D^{(1)}_{m} D^{(1)}_{m'} + D^{(2)}_{m} D^{(2)}_{m'} -2 D^{(3)}_{m} D^{(3)}_{m'} \Big)\\
 &\quad\quad- \Big( D^{(1)}_{m} D^{(3)}_{m'} + D^{(3)}_{m} D^{(1)}_{m'} \Big) \\
 &\quad\quad\quad- \Big( D^{(2)}_{m} D^{(3)}_{m'} + D^{(3)}_{m} D^{(2)}_{m'} \Big) \\
 &\quad\quad\quad\quad-2  \Big( D^{(1)}_{m} D^{(2)}_{m'} + D^{(2)}_{m} D^{(1)}_{m'} \Big)   \Bigg].
\end{align*}
\begin{align*}
D^{[2]}_{L=1,\mathsf{MA}}=&\frac{1}{\sqrt{6}} \big\langle 1m; 1m'\big| 1M\big\rangle \\
&\quad\Bigg[ \Big(-D^{(1)}_{m} D^{(1)}_{m'} + D^{(2)}_{m} D^{(2)}_{m'} \Big)\\
 &\quad\quad+ \Big( D^{(1)}_{m} D^{(3)}_{m'} + D^{(3)}_{m} D^{(1)}_{m'} \Big) \\
 &\quad\quad\quad- \Big( D^{(2)}_{m} D^{(3)}_{m'} + D^{(3)}_{m} D^{(2)}_{m'} \Big)   \Bigg].
\end{align*}
In each case there are terms which are symmetric in $m \leftrightarrow m'$ while $\big\langle 1m; 1m'\big| 1M\big\rangle$ is antisymmetric - since derivative operators acting on \emph{different} quarks commute, these terms are zero. The remaining terms, in which both derivatives act on the same quark field, give color-spatial structures proportional to
\begin{align*}
\mathsf{MS} &\sim	\epsilon_{abc}\, \left( (B_M\psi)_a \psi_b\psi_c + \psi_a (B_M\psi)_b \psi_c - 2 \psi_a \psi_b (B_M\psi)_c \right),\\
\mathsf{MA} &\sim	\epsilon_{abc}\, \left( -(B_M\psi)_a \psi_b\psi_c + \psi_a (B_M\psi)_b \psi_c \right).
\end{align*}
The form of the three-quark color coupling can be exposed by expressing the chromomagnetic field in an adjoint basis, $B_M = B_M^A t^A$, so that
\begin{align*}
(\psi\psi\psi)^A_\mathsf{MS} &\sim	\epsilon_{abc}\, \left( (t^A\psi)_a \psi_b\psi_c + \psi_a (t^A\psi)_b \psi_c - 2 \psi_a \psi_b (t^A\psi)_c \right),\\
(\psi\psi\psi)^A_\mathsf{MA} &\sim	\epsilon_{abc}\, \left( -(t^A\psi)_a \psi_b\psi_c + \psi_a (t^A\psi)_b \psi_c \right),
\end{align*}
which indicates that the three quarks are coupled to a color-octet as either $(\mathbf{3}\otimes\mathbf{3}\to \overline{\mathbf{3}})\otimes\mathbf{3} \to \mathbf{8}\sim \mathsf{MS}$ or $(\mathbf{3}\otimes\mathbf{3}\to \mathbf{6})\otimes\mathbf{3} \to \mathbf{8} \sim \mathsf{MA}$ . The chromomagetic field (a color octet) is coupled in to give an overall color singlet baryon operator, $(\psi\psi\psi)^A B_M^A$. These are the operators which we identify with hybrid baryons.

At first glance it would appear that $D^{[2]}_{L=1,\mathsf{S}}$, as defined in equation (\ref{eq:2DS}) and written out in full in equation (\ref{D2S}) should also have hybrid character owing to the presence of terms featuring $\langle 1m;1m'|1M\rangle D^{(n)}_m D^{(n)}_{m'}$. In fact this operator is zero -  $D^{[2]}_{L=1,\mathsf{S}}$ gives rise to a three-quark color structure  
\begin{equation*}
(\psi\psi\psi)^A_\mathsf{S} \sim	\epsilon_{abc}\, \left( (t^A\psi)_a \psi_b\psi_c + \psi_a (t^A\psi)_b \psi_c + \psi_a \psi_b (t^A\psi)_c \right),
\end{equation*}
which cannot be a color-octet as the antisymmetric combination of three $\mathbf{3}_c$ representations gives a color-singlet. We do not include any operators constructed using $D^{[2]}_{L=1,\mathsf{S}}$ in our basis.

We mention in passing that the antisymmetric combination of two derivatives defined in equation (\ref{eq:2DA}), has structure
\begin{equation*}
D^{[2]}_\mathsf{A} = \tfrac{1}{\sqrt{6}}\left([D^{(1)},D^{(2)}]  + [D^{(2)},D^{(3)}] +[D^{(3)},D^{(1)}] \right),
\end{equation*}
which does not feature the commutator of two gauge-covariant derivatives acting on the {\it same} quark field. The antisymmetric nature is such that only with $L=1$ do we get non-zero operators but these are \emph{not} of hybrid nature; in the Jacobi basis presented in the previous section they transform like $\big\langle 1m;1m'\big| 1 M\big\rangle (\vec{p}_\rho)_{m} (\vec{p}_\lambda)_{m'}$, i.e. they can be viewed as a $P$-wave on each ``oscillator" in the language of the constituent quark model, giving states labeled $\,^{2,4}\!P_\mathsf{A}$. 

The interpretations of the $D^{[2]}$ constructions are summarised in Table \ref{d2}. These operators, \emph{excluding} those identified above as being of hybrid character, were used in \cite{Edwards:2011jj} to extract a spectrum of baryons. In the next section of this paper we will present spectrum results including the hybrid operators, identifying a number of states with large overlap onto these gluonic operators that we interpret as being hybrid baryons.
\begin{table}%[h]
\begin{tabular}{r|c|c}
& $L=0,2$ & $L=1$ \\
\hline
$D^{[2]}_\mathsf{S}$ & $(qqq)_{\mathbf{1}_c}$ & 0\\
$D^{[2]}_\mathsf{M}$ & $(qqq)_{\mathbf{1}_c}$ & $\big[(qqq)_{\mathbf{8}_c}G_{\mathbf{8}_c}\big]_{\mathbf{1}_c}$ \\
$D^{[2]}_\mathsf{A}$ & 0 & $(qqq)_{\mathbf{1}_c}$
\end{tabular}
\caption{Summary of simplest bound-state interpretations of two-derivative structures. In practice the derivatives are actually implemented as finite differences through gauge-covariant parallel transport (links) and this gives rise to corrections to the above assignments that are suppressed by powers of the lattice spacing. 
\label{d2}}
\end{table}

We have determined that operators featuring $D^{[2]}_{L=1,\mathsf{M}}$ can be associated with hybrid structure - now we can ask what flavor-spin constructions can be combined with $D^{[2]}_{L=1,\mathsf{M}}$ to give operators which respect Fermi antisymmetry? As is the case with $qqq$ interpretations, the easiest hybrid interpolators to interpret phenomenologically are those with ``non-relativistic" Dirac spin structure. The explicit flavor-spin-color structure of these operators, featuring $D^{[2]}_{L=1,\mathsf{M}}$, is as follows:
\begin{widetext}
\begin{eqnarray}
^2\mathbf{1} :& \left(\Lambda_{\mathbf{1},\mathsf{A}} \otimes \left(\tfrac{1}{2}^+ \right)_\mathsf{M}^\mathit{1} \otimes D^{[2]}_{L=1,\mathsf{M}}\right)^{J^P=\tfrac{1}{2}^+,\tfrac{3}{2}^+} &\sim \phi_\mathsf{A} \tfrac{1}{\sqrt{2}} \left( \psi_{\bar{\mathbf{3}}} \chi_\mathsf{MA} + \psi_\mathbf{6} \chi_\mathsf{MS} \right) = \phi_\mathsf{A} \left(\psi \chi \right)_\mathsf{S}= \left[\psi\phi\chi \right]_\mathsf{A}  \nonumber \\
%%%%%%%%%%%%%%%%%%%%
^2\mathbf{8} :& \left(\left\{N,\Sigma_\mathbf{8}, \Lambda_\mathbf{8},\Xi_\mathbf{8} \right\}_\mathsf{M} \otimes \left(\tfrac{1}{2}^+ \right)_\mathsf{M}^\mathit{1} \otimes D^{[2]}_{L=1,\mathsf{M}}\right)^{J^P=\tfrac{1}{2}^+,\tfrac{3}{2}^+} &\sim \tfrac{1}{2}\left[ \psi_{\bar{\mathbf{3}}} \left( \phi_\mathsf{MA} \chi_\mathsf{MA}- \phi_\mathsf{MS} \chi_\mathsf{MS}  \right) - \psi_{\mathbf{6}} \left( \phi_\mathsf{MA} \chi_\mathsf{MS} + \phi_\mathsf{MS} \chi_\mathsf{MA}  \right) \right] \nonumber \\
&&= \tfrac{1}{\sqrt{2}} \left[ \psi_{\bar{\mathbf{3}}} \left(\phi\chi \right)_\mathsf{MS} - \psi_\mathbf{6} \left(\phi\chi \right)_\mathsf{MA}  \right] = \left[\psi\phi\chi \right]_\mathsf{A} \nonumber\\
%%%%%%%%%%%%%%%%%%%%%%
^4\mathbf{8} :& \left(\left\{N,\Sigma_\mathbf{8}, \Lambda_\mathbf{8},\Xi_\mathbf{8} \right\}_\mathsf{M} \otimes \left(\tfrac{3}{2}^+ \right)_\mathsf{S}^\mathit{1} \otimes D^{[2]}_{L=1,\mathsf{M}}\right)^{J^P=\tfrac{1}{2}^+,\tfrac{3}{2}^+,\tfrac{5}{2}^+ } &\sim \tfrac{1}{\sqrt{2}} \chi_\mathsf{S} \left( \psi_{\bar{\mathbf{3}}}  \phi_\mathsf{MS} - \psi_{\mathbf{6}}  \phi_\mathsf{MA}   \right)  = \chi_\mathsf{S} \left( \psi \phi \right)_\mathsf{A} = \left[\psi\phi\chi \right]_\mathsf{A} \nonumber\\
%%%%%%%%%%%%%%%%%%%%%%%
^2\mathbf{10} :& \left(\left\{\Delta,\Sigma_\mathbf{10},\Xi_\mathbf{10},\Omega \right\}_\mathsf{S} \otimes \left(\tfrac{1}{2}^+ \right)_\mathsf{M}^\mathit{1} \otimes D^{[2]}_{L=1,\mathsf{M}}\right)^{J^P=\tfrac{1}{2}^+,\tfrac{3}{2}^+} &\sim \tfrac{1}{\sqrt{2}} \phi_\mathsf{S} \left( \psi_{\bar{\mathbf{3}}}  \chi_\mathsf{MS} - \psi_{\mathbf{6}}  \chi_\mathsf{MA}   \right)  = \phi_\mathsf{S} \left( \psi \chi \right)_\mathsf{A} = \left[\psi\phi\chi \right]_\mathsf{A} \label{structures}
\end{eqnarray}
\end{widetext}
where $\psi_{\bar{\mathbf{3}}},\, \psi_{\mathbf{6}}$ are the $\mathsf{MA},\, \mathsf{MS}$ color combinations (corresponding to the $\mathsf{MS},\, \mathsf{MA}$ derivative combinations) and $\phi, \chi$ are the flavor, spin structures presented in Appendix A of \cite{Edwards:2011jj}. This set of operators corresponds to a $[\mathbf{70},1^+]$ in the $SU(6)_{FS}$ assignment scheme, a representation not present for conventional three-quark baryons.

Coupling the hybrid derivative structures ($D^{[2]}_{L=1,\mathsf{M}}$) to Dirac spin lower components gives a somewhat larger set of operators, including some with negative parity, as indicated in Table \ref{hyb_tab}, although their model interpretation is somewhat more involved and will not be described here.

\begin{table*}
\begin{tabular}{rr|rr|rr|rr}
\multicolumn{4}{c|}{positive parity} & \multicolumn{4}{|c}{negative parity} \\
\hline
$N(\tfrac{1}{2}^+)$ & 32\{4\}[8](2) & $\Delta(\tfrac{1}{2}^+)$ & 16\{2\}[4](1) 
& $N(\tfrac{1}{2}^-)$ & 32\{2\}[8](0) & $\Delta(\tfrac{1}{2}^-)$ & 16\{1\}[4](0) \\

$N(\tfrac{3}{2}^+)$ & 36\{5\}[8](2) & $\Delta(\tfrac{3}{2}^+)$ & 19\{3\}[4](1) 
& $N(\tfrac{3}{2}^-)$ & 36\{2\}[8](0) & $\Delta(\tfrac{3}{2}^-)$ & 19\{1\}[4](0) \\

$N(\tfrac{5}{2}^+)$ & 19\{3\}[3](1) & $\Delta(\tfrac{5}{2}^+)$ & 9\{2\}[1](0) 
& $N(\tfrac{5}{2}^-)$ & 19\{1\}[3](0) & $\Delta(\tfrac{5}{2}^-)$ & 9\{0\}[1](0) \\

$N(\tfrac{7}{2}^+)$ & 4\{1\}[0](0) & $\Delta(\tfrac{7}{2}^+)$ & 3\{1\}[0](0) 
& $N(\tfrac{7}{2}^-)$ & 4\{0\}[0](0) & $\Delta(\tfrac{7}{2}^-)$ & 3\{0\}[0](0) \\

\end{tabular}
\caption{Number of operators constructed in each quantum number channel: total\{non-relativistic conventional\}[hybrid](non-relativistic hybrid) \label{hyb_tab}}
\end{table*}

In order to account for the reduced (cubic) rotational symmetry of the lattice, we \emph{subduce} all interpolators into irreducible representations of the cubic group and form correlators according to that symmetry. In \cite{Edwards:2011jj} we show that the spectrum of states extracted from variational analysis of these correlators can be spin-identified with confidence and herein we will present the spectra labelled according to the continuum spin.

%%%%%%%%%%%%%%%%%%%%%%%%%%%%%%%%%%%%%%%%%%%%%%%%
%%%%%				RESULTS  				%%%%%
%%%%%%%%%%%%%%%%%%%%%%%%%%%%%%%%%%%%%%%%%%%%%%%%

\section{Baryon Spectrum}\label{sec:results}

Mass spectra and matrix-elements follow from variational analysis of a matrix of correlators, $C_{ij}(t) = \big\langle 0 \big | \mathcal{O}_i(t) \mathcal{O}^\dag_j(0) \big| 0 \big\rangle$, which can be shown to correspond to solution of the generalised eigenvalue problem
\begin{equation*}
	C(t) v_\mathfrak{n} = \lambda_\mathfrak{n}(t,t_0) C(t_0) v_\mathfrak{n}.
\end{equation*}
Each state $\big|\mathfrak{n}\big\rangle$ that can be interpolated by some linear combination of $\mathcal{O}_i^\dag$ has associated with it a ``principal correlator", $\lambda_\mathfrak{n}(t,t_0)$ that at large times behaves like $e^{-E_\mathfrak{n} (t-t_0)}$, yielding the energy of the state, and an eigenvector $v_\mathfrak{n}$ which indicates the optimum interpolator for the state, $\sum_i v_\mathfrak{n}^i \mathcal{O}^\dag_i$. Details of our procedure for efficiently and accurately implementing such a solution can be found in \cite{Dudek:2007wv,Dudek:2010wm}.

To the extent that one has a bound-state interpretation of the interpolator $\mathcal{O}_i$, built from quark and gluon fields, the relative sizes of the matrix elements $\big\langle \mathfrak{n} \big| \mathcal{O}^\dag_i(0) \big| 0\big\rangle$ can be used to interpret the bound-state structure of state $\big| \mathfrak{n}\big\rangle$. These matrix elements are trivially related to the eigenvectors of the generalised eigenvalue problem solved above. Such a procedure was followed in \cite{Dudek:2010wm}, leading to identification of the low-lying meson spectrum as being rather like the constituent quark model $q\bar{q}$ spectrum, supplemented with a spectrum of hybrid mesons in which a chromomagnetic gluonic field was dominant. We will take the same approach here in the baryon spectrum in order to isolate the role of hybrid baryon basis states in the spectrum.

We performed calculations on lattices of size $16^3\times128$, corresponding to a physical spatial volume of approximately $(2.0\,\mathrm{fm})^3$, with three flavors of dynamical quarks. One quark has mass corresponding to the strange quark and the other two correspond to degenerate $u$ and $d$ quarks. One calculation had all three quarks degenerate at the strange quark mass, corresponding to 702 MeV octet pseudoscalar bosons, and the remaining two calculations had lighter $u,d$ quarks giving pion masses of 524 and 396 MeV. Details of these dynamical anisotropic clover configurations can be found in \cite{Edwards:2008ja,Lin:2008pr}. The two-point correlators used in the variational analysis were computed via distillation\cite{Peardon:2009gh} in which all quark fields are smeared over space by an operator $\Box = \sum_n^N \xi_n \xi^\dag_n$ constructed using the lowest $N=56$ eigenvectors of the three-dimensional gauge-covariant laplacian ($-\vec{D}^2 \xi_n = \lambda_n \xi_n$). All gauge-links entering in the operator constructions are stout-smeared \cite{Morningstar:2003gk}.

To facilitate comparisons of the spectrum at different quark masses, the ratio of hadron masses with the $\Omega$ baryon mass obtained on the same ensemble is used to remove the explicit scale dependence, following Ref.~\cite{Lin:2008pr}. 

\subsection{Nucleons \& Deltas at $m_\pi = 524$ MeV}

\begin{figure*}
\includegraphics[width=.85\textwidth%, bb=0 30 576 360
]{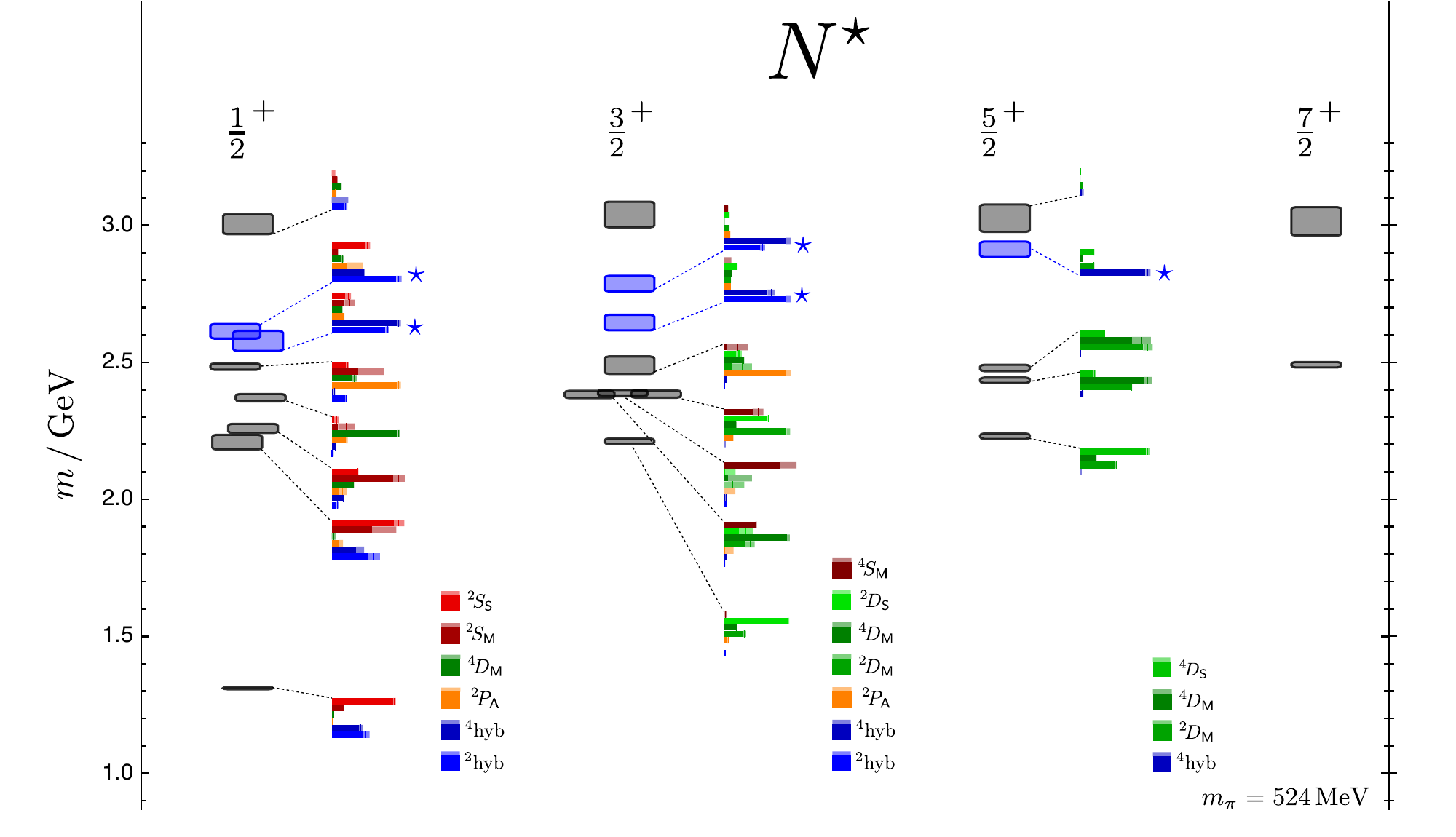}

\vspace{.5cm}

\includegraphics[width=.85\textwidth%, bb=0 30 576 360
]{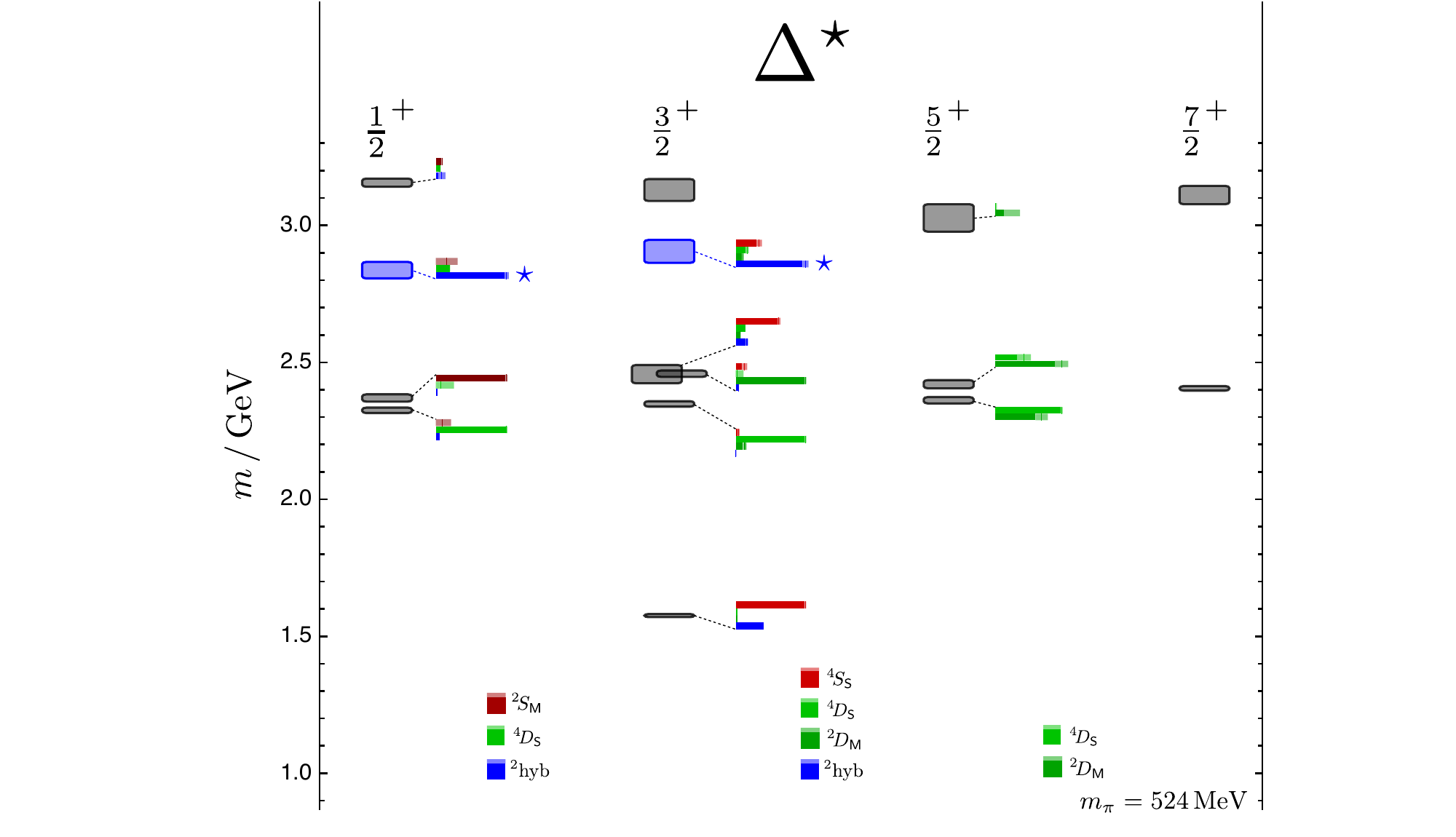}
\caption{
Extracted spectrum of Nucleon and Deltas states by $J^P$ at a pion mass of 524 MeV. Rectangle height indicates the statistical uncertainty in mass determination. Histograms indicate the relative size of matrix elements $\big\langle \mathfrak{n} \big| \mathcal{O}^\dag_i(0) \big| 0\big\rangle$ for a ``non-relativistic" subset of the operators used - normalisation is as described in the text with the lighter area at the head of each bar being the statistical uncertainty. Operator labeling is as in Table IV of \cite{Edwards:2011jj} with the addition of $^2\mathrm{hyb} = \big(N_\mathsf{M} \otimes \big(\tfrac{1}{2}^+\big)^\mathit{1}_\mathsf{M} \otimes D^{[2]}_{L=1,\mathsf{M}}\big)$ and $^4\mathrm{hyb} = \big(N_\mathsf{M} \otimes \big(\tfrac{3}{2}^+\big)^\mathit{1}_\mathsf{S} \otimes D^{[2]}_{L=1,\mathsf{M}}\big)$. Asterisks indicate states having dominant overlap onto hybrid operators.
\label{808histo}}
\end{figure*}

Figure \ref{808histo} shows the spectrum of positive parity Nucleon and Delta states extracted on the 524 MeV lattice along with histograms showing the relative sizes of matrix elements $\big\langle \mathfrak{n} \big| \mathcal{O}^\dag_i(0) \big| 0\big\rangle$ for a set of ``non-relativistic" interpolators characteristic of $qqq$ and hybrid basis states. The histograms are normalised such that the largest overlap onto a given operator across the entire extracted spectrum has unit value (the tallest bars in the figure are of unit value).

%%%%%%%%%%%%%%%%%%%%%%%%%%%%%%%%%%%%
%%%%%%%%%%%%%%%%%%%%%%%%%%%%%%%%%%%%%

Focussing first on the $J^P=\tfrac{1}{2}^+$ nucleon states we observe a light ground state nucleon having dominant\footnote{
The observed sub-dominant overlap of the ground state nucleon onto operators featuring an excited gluonic field is also required to describe QCD sum-rules of the $N_{1/2^+}$ channel \cite{Kisslinger:1995yw}
} overlap onto a ``non-relativistic" interpolator $\big(N_\mathsf{M} \otimes \big(\tfrac{1}{2}^+\big)^\mathit{1}_\mathsf{M} \otimes D^{[2]}_{L=0,\mathsf{S}}\big)^{\tfrac{1}{2}^+}$ with $qqq$ structure $^2\!S_\mathsf{S}$. The band of four excited $\frac{1}{2}^+$ states between 2.0 and 2.5 GeV can be characterised as being dominantly admixtures of the $qqq$ basis $^2\!S_\mathsf{S},\, ^2\!S_\mathsf{M}, \, ^4\!D_\mathsf{M},\, ^2\!P_\mathsf{A}$, and while this basis is clearly not manifested diagonally in the spectrum, the mixing is modest. 
Notably, the extracted spectrum lacks a light first-excited state that we might associate with the broad enhancement observed experimentally and known as the ``Roper" resonance. We will return to this later when we discuss the role of hadronic decays on the spectrum.

Slightly heavier than the first-excited band of four $\tfrac{1}{2}^+$ states we observe two states with dominant overlap onto non-relativistic interpolators we have identified as being of hybrid nature, $\big(N_\mathsf{M} \otimes \big(\tfrac{1}{2}^+\big)^\mathit{1}_\mathsf{M} \otimes D^{[2]}_{L=1,\mathsf{M}}\big)^{\tfrac{1}{2}^+}$ and $\big(N_\mathsf{M} \otimes \big(\tfrac{3}{2}^+\big)^\mathit{1}_\mathsf{S} \otimes D^{[2]}_{L=1,\mathsf{M}}\big)^{\tfrac{1}{2}^+}$. They appear to be largely admixtures of two basis states with three-quark spin $S=1/2,\,3/2$ that we label $^2\mathrm{hyb},\, ^4\mathrm{hyb}$.
We mention in passing that in \cite{Edwards:2011jj}, the spectrum extracted {\it without} hybrid interpolators showed one badly-determined state in this same mass region, indicating the importance of including interpolators with good overlap onto all types of state present in the spectrum.

The next nucleon state resolved with $J^P=\tfrac{1}{2}^+$ is somewhat heavier than the hybrids and has no obvious interpretation in terms of overlaps with the ``non-relativistic" operators used (with up to two gauge-covariant derivatives). Still heavier states are present with these quantum numbers\footnote{There are a total of 32 $\tfrac{1}{2}^+$ states extracted, equal to the number of interpolators used, but the statistical precision and stability of extraction of the masses decreases as one goes up in energy}. 

In the $J^P=\tfrac{3}{2}^+$ nucleon channel, we see a lightest band of states roughly degenerate with the first-excited band in $\tfrac{1}{2}^+$. This band features five states observed to be dominated by the $qqq$ basis $^4\!S_\mathsf{M},\, ^2\!D_\mathsf{S},\, ^4\!D_\mathsf{M},\, ^2\!D_\mathsf{M},\, ^2\!P_\mathsf{A}$, with the state dominated by $^2\!P_\mathsf{A}$ being heaviest as it was for $J^P=\tfrac{1}{2}^+$.
As in the nucleon $\tfrac{1}{2}^+$ case we observe a pair of states somewhat above this band which have dominant overlap onto $^2\mathrm{hyb},\, ^4\mathrm{hyb}$. A few hundred MeV heavier there are more states of unidentified structure.

In the $J^P=\tfrac{5}{2}^+$ nucleon channel there are three states in the ``first-excited" mass region, apparently constructed from superpositions of $^4\!D_\mathsf{S},\, ^2\!D_\mathsf{M},\, ^4\!D_\mathsf{M}$. Somewhat heavier is a single state having large overlap onto a hybrid operator of character $^4\mathrm{hyb}$. Unidentified $\tfrac{5}{2}^+$ states are present a little heavier than the hybrid.

In the $J^P=\tfrac{7}{2}^+$ nucleon channel, there is a single state in the ``first-excited" region having dominant overlap onto $^4\!D_\mathsf{M}$. Heavier states of unidentified structure are present above 3 GeV, with no identified hybrid states. We remind the reader that there are no ``non-relativistic" hybrid interpolators having $J^P=\tfrac{7}{2}^+$ constructed at $D^{[2]}$ level. 

In the Delta sector we again see a band of ``first-excited" states whose population and overlaps are in accord with expectations of a $qqq$ model. States with dominant overlap onto our hybrid interpolators appear above that band, but below other unidentified states. The hybrids appear in the $J^P=\tfrac{1}{2}^+$, $\tfrac{3}{2}^+$ channels and are of $^2\mathrm{hyb}$ character.

The spectrum of negative parity nucleons and Deltas below 2 GeV is almost identical to that presented in \cite{Edwards:2011jj}, obtained without inclusion of any hybrid interpolators, and agrees with $qqq$ model state counting. Negative parity states having significant overlap onto the hybrid operators in the basis first appear above 3 GeV, somewhat heavier than the positive parity hybrid states.

\subsection{$\mathbf{1}_F,\,\mathbf{8}_F,\,\mathbf{10}_F$ at an $SU(3)_F$ symmetric point}

\begin{figure*}
 \centering
\includegraphics[width=0.79\textwidth%, bb=0 30 576 360
]{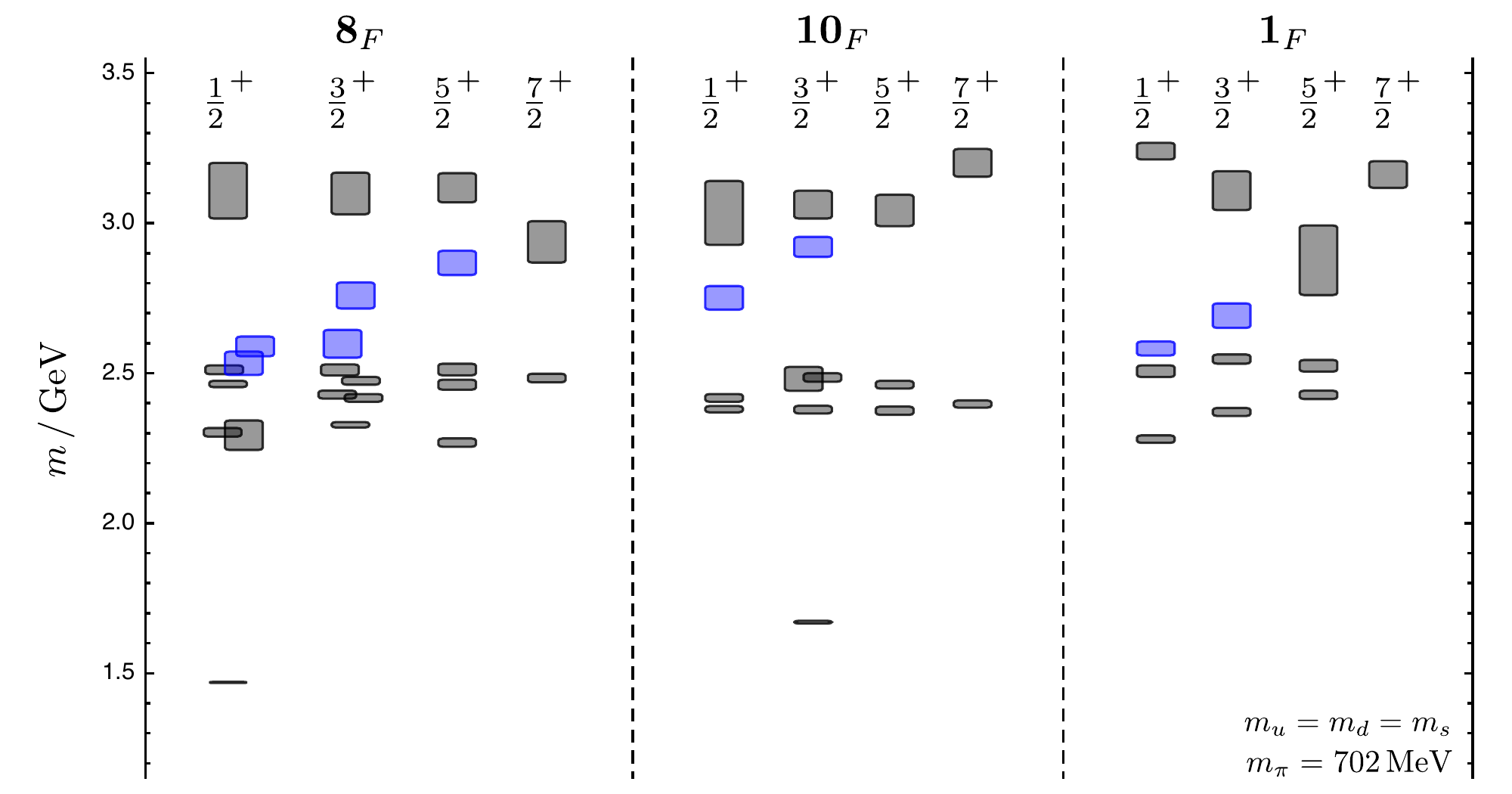}
\caption{Extracted spectrum of flavor octet, decuplet and singlet states by $J^P$ with three quark flavors all at the strange quark mass, corresponding to an octet pseudoscalar mass of 702 MeV. Grey boxes are conventional $qqq$ states and blue boxes are the states identified as hybrid baryons.\label{743}}
\end{figure*}

The flavor singlet construction in equation (\ref{structures}), $\Lambda_{\mathbf{1},\mathsf{A}} \otimes \big(\tfrac{1}{2}^+\big)^\mathit{1}_\mathsf{M} \otimes D^{[2]}_{L=1,\mathsf{M}}$, is most simply explored at an $SU(3)_F$ point where the states interpolated cannot mix with Lambda states in an $\mathbf{8}_F$ representation. In Figure \ref{743} we show the spectrum extracted with all three quark masses set to the strange quark mass - here the lightest pseudoscalar meson has a mass of 702 MeV. We show the mass spectrum for $\mathbf{8}_F,\, \mathbf{10}_F,\, \mathbf{1}_F$ representations.

We observe that the $\mathbf{8}_F,\, \mathbf{10}_F$ spectra strongly resemble the Nucleon, Delta spectra extracted at the 524 MeV pion mass shown in figure \ref{808histo}. The $\mathbf{1}_F$ spectrum consists of the states expected in a $qqq$ model, lying roughly degenerate with the first excited band in the octet, decuplet spectrum, and in addition, two states of hybrid character having $J^P=\tfrac{1}{2}^+,\, \tfrac{3}{2}^+$. As in the octet and decuplet, the hybrid states are slightly heavier than the $qqq$ states.

\subsection{Quark mass dependence}

\begin{figure}[h]
 \centering
\includegraphics[width=0.49\textwidth%, bb=0 30 576 360
]{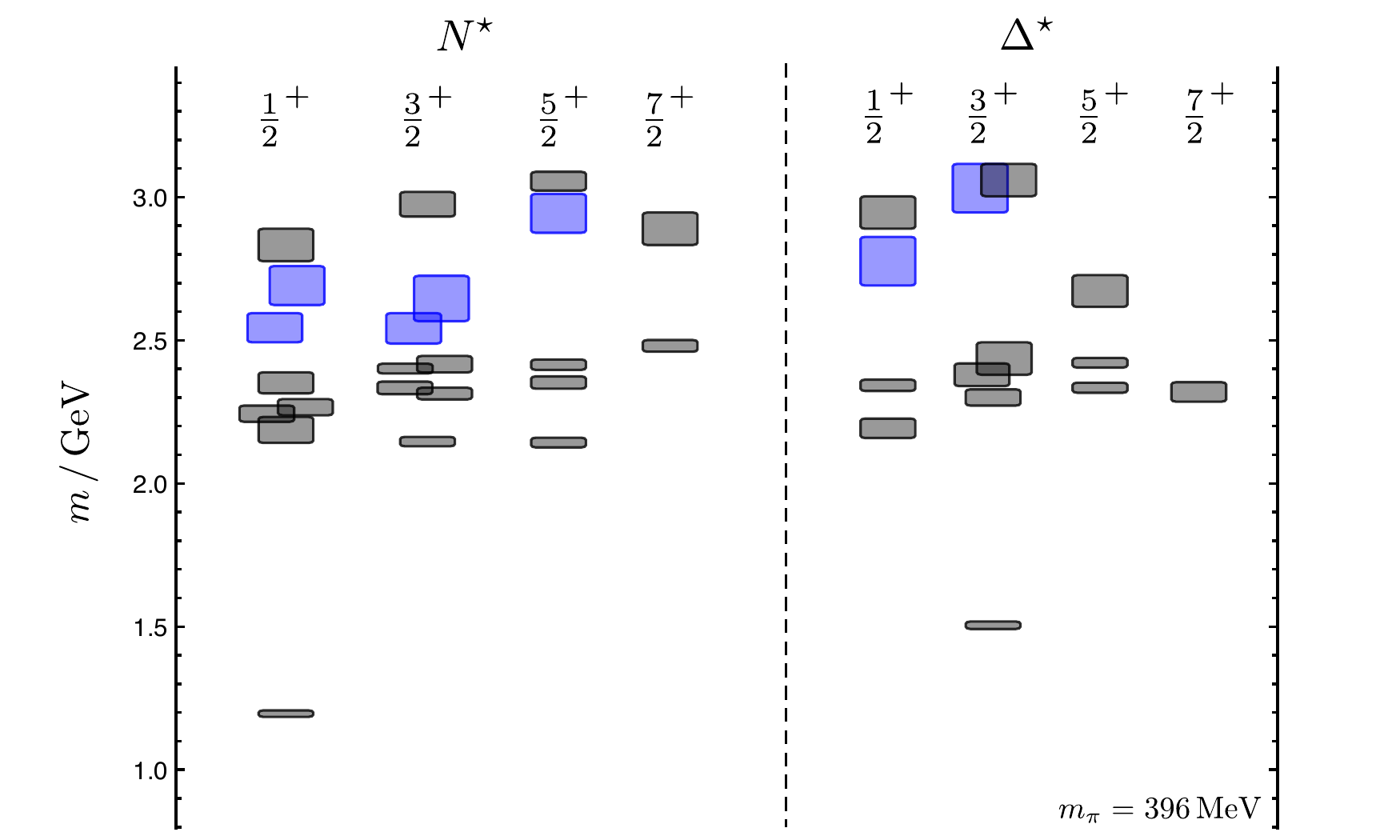}
\caption{Extracted spectrum of Nucleon and Delta states at a pion mass of 396 MeV. Grey boxes are conventional $qqq$ states and blue boxes are the states identified as hybrid baryons.\label{840}}
\end{figure}

In order to determine if the hybrid spectrum might be strongly quark mass dependent we also computed the Nucleon and Delta spectrum with $u,d$ quark masses such that the pion weighs 396 MeV. The results are presented in Figure \ref{840}, where we see that the gross structure of the spectrum is as it was at $m_\pi = 524 \,\mathrm{MeV}$.

%%%%%%%%%%%%%%%%%%%%%%%%%%%%%%%%%%%%%%%%%
%%   DISCUSSING THE RESULTS
%%%%%%%%%%%%%%%%%%%%%%%%%%%%%%%%%%%%%%%%%

\section{Conclusions}\label{sec:conc}

From the extracted baryon spectra presented in the previous section we can draw a number of conclusions regarding the nature of hybrid baryons and more generally gluonic excitations within QCD. The states identified have large overlap onto interpolating fields containing the operator $D^{[2]}_{L=1,\mathsf{M}}$ which transforms like a chromomagnetic field (color octet, $J^{PC}=1^{+-}$)\footnote{see \cite{Dudek:2011bn} for a simple model in which this operator interpolates a quasi-gluon in a $P$-wave}. The form of the operators suggest that within a hybrid baryon the three quarks are arranged in a color octet with the chromomagnetic gluonic excitation making the state an overall color singlet. The low-lying hybrid states overlap strongly onto the ``non-relativistic" subset of our hybrid interpolators, those constructed using only upper components of Dirac spinors. We can interpret this as suggesting that the quarks within the lightest hybrid baryons are dominantly in $S$-waves. 

The particular set of flavor-$J^P$ states observed, $N_{1/2^+},\,N_{1/2^+},\,N_{3/2^+},\, N_{3/2^+},\,N_{5/2^+},\,$ plus $\Delta_{1/2^+},\, \Delta_{3/2^+}$ and $\Lambda_{\mathbf{1},1/2^+},\, \Lambda_{\mathbf{1},3/2^+}$, suggests that all color singlet states are present that can be constructed from antisymmetric $(qqq)_{\mathbf{8}_c}$ with quarks in an $S$-wave coupled to a chromomagnetic ($\mathbf{8}_c$, $1^+$) gluonic excitation (see equations (\ref{structures})). Furthermore the fact that we are easily able to pick the hybrid states out of a dense spectrum of $qqq$ states indicates that they tend to retain their character despite the possibility of mixing strongly with $qqq$ states of the same $J^P$.

The position of hybrid baryons within the spectrum of expected $qqq$ states is now determined: the lightest states are of positive parity and lie slightly heavier than the first-excited positive parity $qqq$ states. Negative parity hybrid baryons appear to be heavier than this.
 
\begin{figure*}
 \centering
\includegraphics[width=0.99\textwidth%, bb=0 30 576 360
]{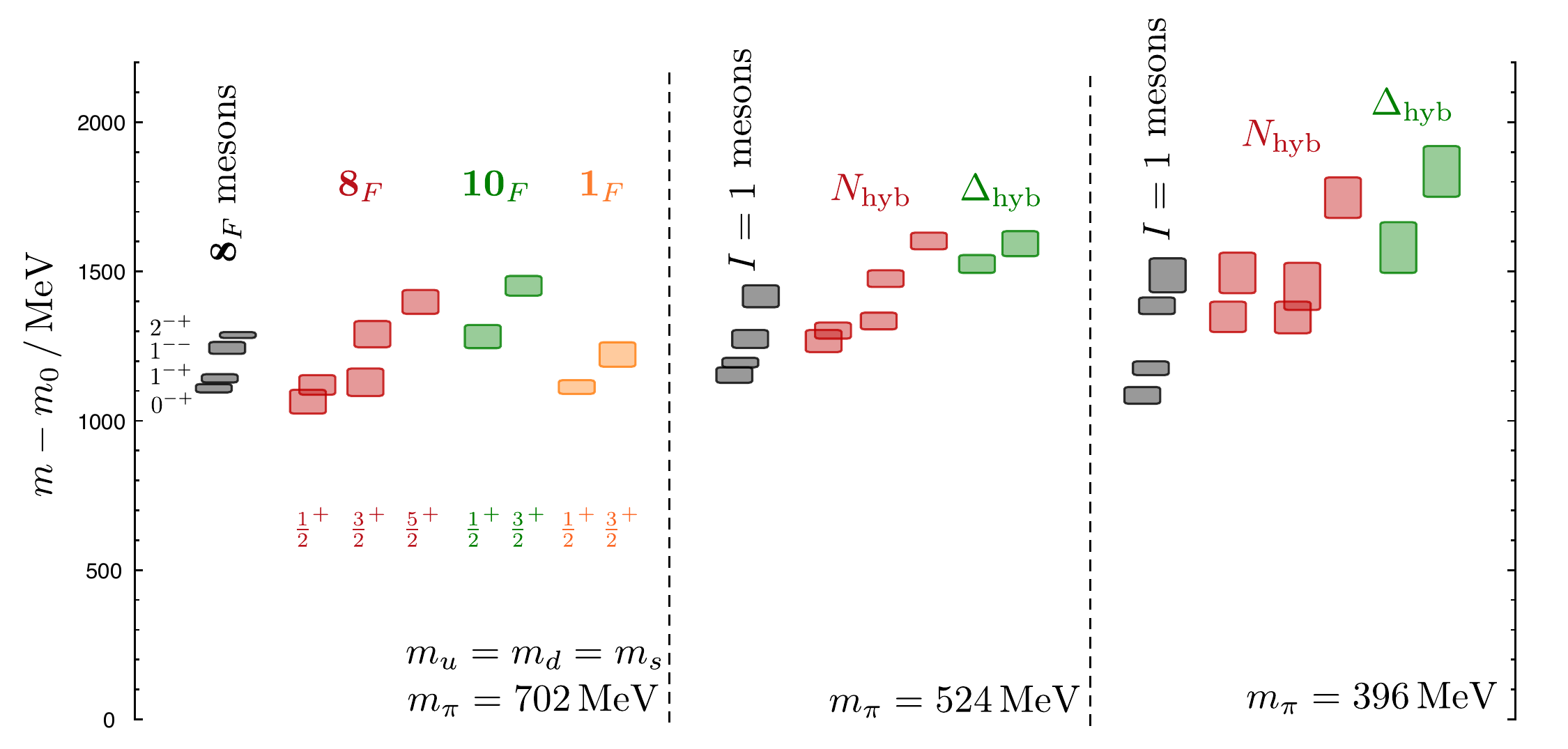}
\caption{Spectrum of hybrid mesons and baryons for three light quark masses. Mass scale is $m-m_\rho$ for mesons and $m-m_N$ for baryons to approximately subtract the effect of differing numbers of quarks. \label{mesons}}
\end{figure*}

In \cite{Dudek:2011bn} we presented the spectrum of hybrid mesons extracted on the same lattices. The distribution of states across $J^{PC}$ (the supermultiplet structure) is also compatible with a color octet chromomagnetic field coupled to quarks in a color octet (in this case $(\bar{q}q)_{\mathbf{8}_c}$). This leads us to think that the gluonic excitation form may be common to hybrid mesons and baryons. To determine whether the energy scale of the gluonic excitation is common we choose to plot the spectrum of hybrid mesons alongside the spectrum of hybrid baryons. Of course for this to be meaningful we must take some account of the differing number of quarks - in a constituent quark model we would subtract twice the constituent quark mass in mesons and three times the mass in baryons. With the lattice data we opt to subtract the $\rho$ mass from the meson spectrum and the nucleon mass from the baryon spectrum. This is presented in Figure \ref{mesons} where it would appear that there is a common energy scale, in the region of 1.3 GeV, for the lowest gluonic excitation in mesons and baryons. 
 
It is worth mentioning here that in the meson sector we used an operator basis featuring up to three derivatives. The low-lying spectrum of negative-parity hybrid mesons was largely insensitive to the presence or absence of the three-derivative operators. Conversely, accurate determination of the higher-lying positive-parity hybrid mesons required inclusion of these operators. We expect that a similar situation would hold true in the baryon spectrum - where we to include a set of three-derivative operators, the low-lying positive parity hybrid baryon spectrum would be largely unchanged, but we would likely resolve more clearly a spectrum of higher-lying negative parity hybrid baryons.

There are features of our extracted spectrum that we cannot explain without more detailed modelling, in particular the nature of the dynamics giving rise to the fine structure in the spectrum is not known to us. It is interesting to note that the extracted structure is rather close to that presented in \cite{Barnes:1982fj} following from perturbative corrections (quark-spin, gluon Compton etc ...) to the bag spectrum. One possible exercise using the spectrum obtained in this paper would be to explore the minimal set of model components required to describe the observed fine structure.

With optimum linear combinations of baryon interpolators determined in the variational solution of the matrix of two-point correlators, we can in the near future consider computations of radiative transition form-factors. Such calculations, which involve three-point hadron correlators, have been done for charmonium \cite{Dudek:2009kk}, and for spin-$1/2$ baryons, albeit with very small operator bases \cite{Lin:2008qv, Lin:2011da}. Radiative transition form-factors are of phenomenological interest because they can be measured in baryon electroproduction experiments\cite{Mokeev:2011ic} where the photon $Q^2$ dependence and the relative size of multipole amplitudes might provide some insight into the internal structure of the excited state \cite{Li:1991yba}.

The physics of excited states as resonances, able to decay into meson-baryon final states is incomplete in this study. Along with larger than physical quark masses, this may be the cause of the lack of a light Roper candidate state within the calculation. In order to observe such decay physics in the extracted finite volume spectrum, which would correspond to additional volume-dependent energy levels, we must supplement our basis of interpolating fields with some resembling multi-hadron constructions (e.g. $\pi N$, $\pi\pi N$). Such work is underway and may eventually lead to phenomenological decay filters for hybrid character that would be useful for experimental searches.

A more complete understanding of hybrid baryons in QCD will of course require calculations at lighter quark masses, approaching the true physical values, coupled with larger lattice volumes to accommodate the increasing physical size of the bound states and such work is now warranted given the success of this first calculation.

In summary we have extracted a spectrum of baryons from QCD computations that features `expected' $qqq$ states built from superpositions of flavor-spin representations $[\mathbf{56}, 0^+]$, $[\mathbf{70}, 0^+]$, $[\mathbf{70}, 2^+]$, $[\mathbf{56}, 2^+]$, $[\mathbf{20}, 1^+]$ in positive parity and $[\mathbf{70}, 1^-]$ in negative parity, supplemented with a set of positive parity hybrid baryons that form a $[\mathbf{70}, 1^+]$ and which lie above the band of first-excited positive parity baryons. In light of this it seems unlikely that the Roper is dominantly of hybrid character as has been speculated in the past. The structure of the operators that interpolate the hybrid states from the vacuum, along with the observation that the energy scale of gluonic excitations appears to be common for mesons and baryons, provides evidence that the gluonic excitation sector in QCD may turn out to be relatively simple. % and amenable to modeling. 
We suggest that a chromomagnetic excitation ($J^{PC}=1^{+-}$) is lightest with an energy scale in the region of 1.3 GeV.

\acknowledgements

We thank Ted Barnes, Simon Capstick, and Frank Close for their input on the physics of hybrid baryons and our colleagues within the Hadron Spectrum Collaboration. {\tt Chroma}~\cite{Edwards:2004sx} and {\tt QUDA}~\cite{Clark:2009wm,Babich:2010mu} were used to perform this work on clusters at Jefferson Laboratory under the USQCD Initiative and the LQCD ARRA project. Gauge configurations were generated using resources awarded from the U.S. Department of Energy INCITE program at Oak Ridge National Lab, the NSF Teragrid at the Texas Advanced Computer Center and the Pittsburgh Supercomputer Center, as well as at Jefferson Lab. RGE and JJD acknowledge support from U.S. Department of Energy contract DE-AC05-06OR23177, under which Jefferson Science Associates, LLC, manages and operates Jefferson Laboratory. JJD also acknowledges the support of the Jeffress Memorial Fund and the U.S. Department of Energy Early Career award contract DE-SC0006765.

\bibliography{bibliography} 

\end{document}